\documentclass[11pt]{article}

\usepackage{amsmath}
\usepackage{amssymb}

\usepackage{graphicx}

\usepackage{cite}

\usepackage{color} 

\usepackage[labelfont=bf,labelsep=period,justification=raggedright]{caption}

\makeatletter
\renewcommand{\@biblabel}[1]{\quad#1.}
\makeatother

\date{}

\begin{document}

\begin{center}
{\LARGE Radial and spiral stream formation in \textit{Proteus mirabilis}
colonies}

\medskip
{Chuan Xue$^{1,\ast}$, 
Elena O. Budrene$^{2}$, 
Hans G. Othmer$^{3}$}
\end{center}
 
{\small \noindent
$^{1}$ Mathematical Biosciences Institute, the Ohio State University, Columbus, OH, USA \\
$^{2}$ Department of Mathematics, Massachusetts Institute of Technology, Cambridge, MA, USA \\
$^{3}$ School of Mathematics and Digital Technology Center, University of Minnesota, Minneapolis, MN, USA \\
$\ast$ E-mail: cxue@mbi.osu.edu
}
 
\section*{Abstract}
The enteric bacterium \textit{Proteus mirabilis}, which is a pathogen that forms biofilms
\textit{in vivo}, can swarm over hard surfaces and form concentric ring patterns
in colonies. Colony formation involves two distinct cell types: swarmer cells
that dominate near the surface and the leading edge, and swimmer cells that
prefer a less viscous medium, but the mechanisms underlying pattern formation
are not understood. New experimental investigations reported here show that
swimmer cells in the center of the colony stream inward toward the inoculation
site and in the process form many complex patterns, including radial and spiral
streams, in addition to concentric rings. These new observations suggest that
swimmers are motile and that indirect interactions between them are essential in
the pattern formation.  To explain these observations we develop a hybrid
cell-based model that incorporates a chemotactic response of swimmers to a
chemical they produce. The model predicts that formation of radial streams can
be explained as the modulation of the local attractant concentration by the
cells, and that the chirality of the spiral streams can be predicted by
incorporating a swimming bias of the cells near the surface of the
substrate. The spatial patterns generated from the model are in qualitative
agreement with the experimental observations.



\section*{Introduction}

A variety of spatial patterns in growing bacterial colonies are found
in nature and in the lab
\cite{Budrene:1991:CPF,Budrene:1995:DFS,Woodward:1995:STP,Berg:1996:SBM,
Ben-Jacob:1994:GMC,Ben-Jacob:1997:CCM,Bisset:CSM:1976,Rauprich:1996:PPP,Beer:2009:DCB,Wu:2009:PRD}.
When grown on semi-solid agar with succinate or other TCA cycle intermediates,
\textit{Escherichia coli} cells divide, move and selforganize into patterns ranging from
outward-moving rings of high cell density to chevron patterns, depending on the
initial concentration of the nutrient \cite{Budrene:1991:CPF,
Budrene:1995:DFS}. When grown or just placed in static liquids, cells quickly
reorganize into networks of high cell density comprised of bands and/or
aggregates, after exposure to succinate and other compounds. Chemotactic strains
of \textit{Salmonella typhimurium}, a closely-related species, can also form concentric rings and
other complex patterns in similar conditions
\cite{Woodward:1995:STP,Berg:1996:SBM}. It has been shown that pattern formation
in \textit{E. coli}  and \textit{S. typhimurium}  is caused by chemotactic interactions between the
cells and a self-produced attractant \cite{Budrene:1991:CPF,
Budrene:1995:DFS,Woodward:1995:STP}.  The gram-positive bacterium
\textit{Bacillus subtilis} forms patterns ranging from highly branched
fractal-like patterns to compact forms, depending on the agar and nutrient
concentrations \cite{Ben-Jacob:1994:GMC,Ben-Jacob:1997:CCM}. In all these systems
proliferation, metabolism and movement of individual cells, as well as direct
and indirect interactions between cells, are involved in the patterning process,
but how they influence each other and what balances between them lead to the
different types of patterns can best be explored with a mathematical model.
Understanding these balances would advance our understanding of the formation of
more complex biofilms and other multicellular assemblies
\cite{Shapiro:1998:TAB}.

\textit{Proteus mirabilis}  is an enteric gram-negative bacterium that causes urinary
tract infections, kidney stones and other diseases \cite{Zunino:1994:FNP,
Mobley:1995:SPP, Jansen:2003:VPM, Jones:2005:RSF}. \textit{P. mirabilis}  is also known for
spectacular patterns of concentric rings or spirals that form in {\em Proteus}
colonies when grown on hard agar.  Pattern formation by {\em Proteus} was
described over 100 years ago \cite{Hauser:1885:UFB}, and the nature of these
patterns has since been discussed in many publications.

 \textit{P. mirabilis}  cells grown in liquid medium are
vegetative swimmer cells which are 1-2 $\mu$m long, have 1-10 flagella and move using
a ``run-and-tumble strategy'', similar to that used by \textit{E. coli}
 \cite{Berg:1996:SBM}. Swimmers respond chemotactically to several amino acids,
and can adapt perfectly to external signals \cite{Pearson:2008:CGS}. When grown
on hard agar swimmers differentiate into highly motile, hyperflagellated,
multi-nucleated, non-chemotactic swarmer cells that may be as long as 50-100
$\mu$m, and that move coordinately as ``rafts'' in the slime they produce
\cite{Williams:1978:NSP,Fraser:1999:SM}.  During pattern formation on hard
surfaces swarmer cells are found mainly at the leading edge of the colony, while
swimmers dominate in the interior of the colony
\cite{Hauser:1885:UFB,Williams:1978:NSP,Rauprich:1996:PPP,Douglas:DCZ:1976}. More and more effort 
is put into understanding the mechanism of swarming, but to date little is known about how cells 
swarm and how cells undergo transitions between swimmers and swarmers \cite{Williams:1978:NSP,
Fraser:1999:SM, Rather:2005:SCD,Tremblay:2007:SPE,Kearns:2010:FGB,Zhang:2010:CMD,Wu:2011:MRC}, but understanding these processes and how they
affect colonization could lead to improved treatments of the diseases \textit{P. mirabilis} 
can cause.

Traditionally, formation of periodic cell-density patterns in {\em Proteus}
colonies has been interpreted as a result of periodic changes in velocity of the
colony's front, caused by the cyclic process of differentiation and
dedifferentiation of swimmers into swarmers (see \cite{Rauprich:1996:PPP}).
Douglas and Bisset (1976) described a regime for some strains of \textit{P. mirabilis} 
in which swarmers form a continuously moving front, while concentric rings
of high cell density form wel1 behind that front. This suggests that
pattern formation can occur in the absence of cycles of differentiation
and dedifferentiation. The similarity between this mode of pattern formation
and that of {\em Salmonella} led us to ask whether the underlying mechanism
for pattern formation in \textit{P. mirabilis}  might also be chemotactic aggregation of the actively
moving swimmers behind the colony front.

A number of mathematical models of colony front movement have been proposed, and
 in all of them swimmer cells are nonmotile and swarming motility is described
 as a degenerate diffusion, in that swarmers only diffuse when their density
 exceeds a critical value \cite{Esipov:1998:KMP,
 Medvedev:2000:RSP,Czirok:2001:TPS,Ayati:2006:SPM, Ayati:2007:MRC}. The front
 propagation patterns as a function of various parameters in one model are given
 in \cite{Czirok:2001:TPS}.  Although these models can reproduce the colony
 front dynamics, it remains to justify modeling the swarming motility as a
 diffusion process, since it is likely that the cell-substrate interaction is
 important. To replicate a periodically propagating front, Ayati showed that
 swarmers must de-differentiate if and only if they have a certain number of
 nuclei \cite{Ayati:2006:SPM,Ayati:2007:MRC}.  It was shown that this may result
 from diffusion limitations of intracellular chemicals, but biological evidence
 supporting this assumption is lacking, and further investigation is needed to
 understand the mechanism of front propagation.

Here we report new experimental results for a continuously-expanding front and
show that after some period of growth, swimmer cells in the central part of the
colony begin streaming inward and form a number of complex multicellular
structures, including radial and spiral streams as well as concentric
rings. These observations suggest that swimmer cells are also motile and
communication between them may play a crucial role in the formation of the
spatial patterns. However, additional questions raised by the new findings
include: (1) what induces the inward movement of swimmer cells, (2) why they
move in streams, (3) why radial streams quickly evolve into spiral streams, and
(4) quite surprisingly, why all the spirals wind counterclockwise when viewed
from above. To address these questions we developed a hybrid cell-based model in
which swimmer cells communicate by excreting a chemoattractant to which they
also respond. The model has provided biologically-based answers to the questions
above and guided new experiments.  We have also developed a continuum chemotaxis
model for patterning using moment closure methods and perturbation analysis
\cite{Xue:2009:MMT}, and we discuss how the classical models had to be modified
for \textit{Proteus} patterning.

\section*{Results}

\subsection*{Experimental findings}
\label{sec_experiments}
\begin{figure}[!ht]
\begin{center}
\includegraphics[width=4in]{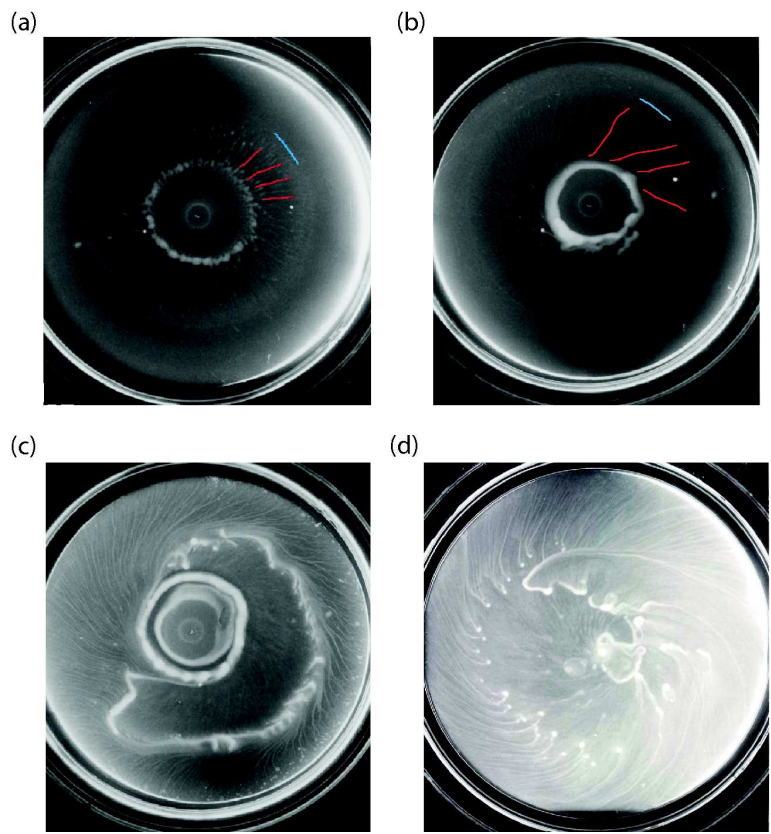}
\end{center}
\caption{ {\bf The evolution of a \textit{P. mirabilis}\, colony.} Time after inoculation: (a) 8.5 hours, (b)
9 hours, and (c) 11 hours. (a) initially homogeneous bacterial lawn breaks into
radial spokes in the central region of the colony, then bacteria and bacterial
aggregates stream inwards following the radial spokes. (b) the radial streams
gradually transform into counterclockwise spirals, and the inner ends of each
arm join together to form a solid toroidal mass.  (c) a second rotating ring
forms with spirals that arise further from the center, and a moving train of high cell
density forms at some distance from the ring. In (a) and (b), the colony front
is highlighted in blue, and a few arms of the streams are highlighted in red. In
(c) the colony has covered the entire  plate. (d) A different experiment that
shows only stream formation without the structure of ring elements.
}\label{fig1}
\end{figure}

\begin{figure}[!ht]
\centering\includegraphics[width=0.4\textwidth]{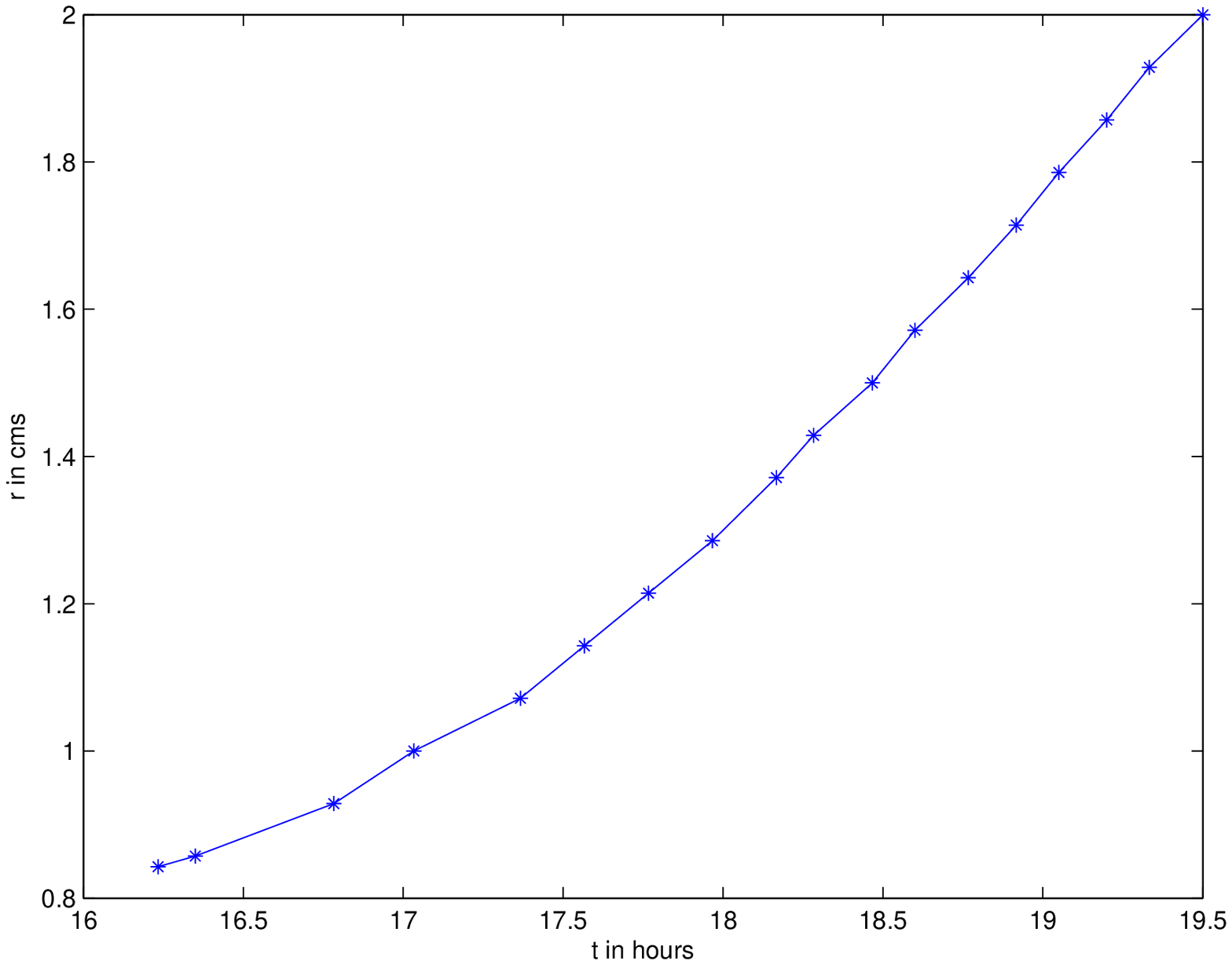}
\caption{The radius of the colony as a function of the real time. Extracted from the supporting movie.}\label{fig_front}
\end{figure}

Previous experimental work focused on expansion of the colony and neglected the
role of swimmers in the pattern formation process. The experimental results
reported here represent a first step toward understanding their role.  After a
drop of \textit{P. mirabilis}  culture is inoculated on a hard agar-like surface containing rich
nutrient, the colony grows and expands.  Under the conditions used here,
the colony front expands continuously initially as a disc of uniform density (Figure \ref{fig_front}).
The swarmers exist
at the periphery of the colony, and the mean length of the cells decreases
towards the center, as observed by others \cite{Matsuyama:2000:DAs}. After a
period of growth, swimmer cells behind the leading edge start to stream inward,
forming a number of complex patterns (Figure \ref{fig1}). The swimmer population
first forms a radial spoke-like pattern in an annular zone on a time scale of
minutes, and then cells follow these radial streams inward (\ref{fig1}a). The
radial streams soon evolve into spirals streams, with aggregates at the inner
end of each arm (\ref{fig1}b). A characteristic feature of this stage is that
the spirals always wind counterclockwise when viewed from above. Different
aggregates may merge, forming more complex attracting structures such as
rotating rings and traveling trains (\ref{fig1}b, c). Eventually the motion
stops and these structures freeze and form the stationary elements of the
pattern (\ref{fig1}b, c). Later, this dynamic process repeats at some distance
from the first element of the pattern, and sometimes cells are recruited from
that element. In this way, additional elements of the permanent pattern are laid
down (\ref{fig1}c). On a microscopic level, transition to the aggregation phase
can be recognized as transformation of a monolayer of cells into a complex
multi-layered structure. Not every pattern is observable in repeated
experiments, (for example, no observable rotating rings can be identified in
(\ref{fig1}d)), probably due to sensitivity to noise in the system and other
factors that require further investigation, variations in nutrient availability,
etc., but the radial and spiral streams seem to be quite reproducible.

These new findings pose challenges to the existing theories of concentric ring
formation in which swimmer cells are believed to be non-motile. Additional
questions arise regarding the mechanism(s) underlying the formation of radial
and spiral streams, rings and trains by swimmers, and what determines the
chirality of the spiral streams.  The macroscopic patterns are very different
and more dynamic than the patterns formed in \textit{Escherichia coli} or 
\textit{Salmonella typhimurium} colonies \cite{Budrene:1991:CPF,Budrene:1995:DFS,Woodward:1995:STP}, where cells interact
indirectly via a secreted attractant, but the fact that swimmers move up the
cell density gradient is quite similar. The non-equilibrium dynamics suggests
intercellular communication between individual swimmer cells. We determined that
swimmer cells extracted from these patterns are chemotactic towards several
amino acids, including Aspartate, Methionine and Serine (see Table 1). In
the following we provide an explanation of the radial and spiral streams using a
hybrid cell-based model.

\subsection*{The hybrid cell-based model}
\label{sec_model} 
\begin{figure}[!ht]
\begin{center}
\includegraphics[width=4in]{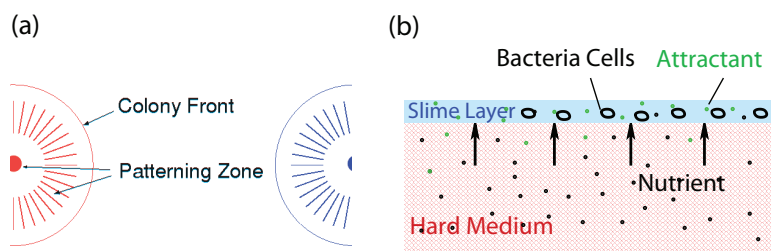}
\end{center}
\caption{ 
{\bf (a) The colony front and the patterning zone.  (b) A
vertical cross-section of the system. }  The lower layer is hard agar that
contains nutrients, and the top layer is slime generated during colony
expansion. Swimmers move in the layer of slime, absorb nutrients that diffuse
upward, and secrete attractant. Bacterial flagella are not shown.}
\label{fig2}
\end{figure}

The spatial patterns
of interest here are formed in the center of the colony where cells are
primarily swimmers, and the role of swarmers is mainly to advance the front and
to affect the swimmer population by differentiation and
de-differentiation. Thus we first focus on modeling the dynamics in the
patterning zone in the colony center (Figure \ref{fig2}a), and later we
incorporate the colony front as a source of swimmers (see Figure \ref{fig6}).
This enables us to avoid unnecessary assumptions on the poorly-understood
biology of swarming and the transition between the two phenotypes.  As noted
earlier, swimmer cells are chemotactic to certain factors in the medium, and we
assume that they communicate via a chemoattractant that they secrete and to
which they respond. Therefore the minimal mathematical model involves equations
for the signal transduction and movement of individual cells, and for the
spatio-temporal evolution of the extracellular attractant and the nutrient in
the domain shown in Figure \ref{fig2}b.  We first focus on understanding the
radial and spiral stream formation, which occurs rapidly, and during which the
nutrient is not depleted and cells grow exponentially. During this period the
nutrient equation is uncoupled from the cell equations and can be ignored.

\textit{P. mirabilis}  is genetically close to \textit{Escherichia coli}, and all the chemotaxis-related genes of
\textit{Escherichia coli} have been identified in \textit{Proteus}
\cite{Pearson:2008:CGS}. \textit{P. mirabilis}  cells swim using a run-and-tumble strategy,
which consists of more-or-less straight runs punctuated by random turns.  In the
absence of an attractant gradient the result is an unbiased random walk, with
mean run time $\sim$ 1 s and mean tumble time $\sim$ 0.1 s. In the presence of
an attractant gradient, runs in a favorable direction are prolonged, and by
ignoring the tumbling time, which is much shorter than the run time, the
movement of each cell can be treated as an independent velocity jump process
with a random turning kernel and a turning rate determined by intracellular
variables that evolve in response to extracellular signals
\cite{Erban:2004:FIC}.  The signal transduction pathway for chemotaxis is
complex and has been studied extensively in  \textit{Escherichia coli},  both experimentally and
mathematically \cite{Lux:2004:CMB,Wadhams:2004:MSI,Bourret:1991:STP,
Spiro:1997:MEA,Shimizu:2003:SES,Mello:2003:QMS,Rao:2004:DDB}.  However the main
processes are relatively simple, and consist of fast excitation in response to
signal changes, followed by adaptation that subtracts out the background
signal.  Given the genetic similarity between \textit{P. mirabilis} and  \textit{Escherichia coli}, we describe
motility and signal transduction of each cell using the key ideas used
successfully for  \textit{Escherichia coli} \cite{ErbanThesis}.

\begin{itemize}
\item Each swimmer cell (with index $i$) is treated as a point and
characterized by its location $\mathbf{x}^{i}$, velocity $\mathbf{v}^{i}$, cell-cycle
clock $A^{i}$ and intracellular variables $\mathbf{y}^{i}$.

\item Signal transduction of each cell is described by the simple
model used in \cite{Erban:2004:FIC}, which captures the main
dynamics of the signal transduction network,
\begin{eqnarray}
\frac{d
y_1^{i}}{dt}&=&\frac{G(S(\mathbf{x},t))-(y_1^{i}+y_2^{i})}{t_e}, \label{internalS}\\
 \frac{d
y_2^{i}}{dt}&=&\frac{G(S(\mathbf{x},t))-y_2^{i}}{t_a}, \label{internalE}
\end{eqnarray}
where $t_e$, $t_a$ with $t_e\ll t_a$ are constants characterizing the excitation and adaptation time
scales, $S$ is the local attractant concentration and $G(S(\mathbf{x},t))$ models detection
and transduction of the signal. Here the variable $y_1$ is the one that excites and adapts to the signal. It has 
a similar role as $\mbox{CheY}_{\mbox{p}}$ in the signal transduction network. The variable $y_2$ causes the adaptation, which 
models the methylation level of receptors.

\item The turning rate and turning kernel are
\begin{equation}
\lambda^{i} =  \lambda_0 \left(1 - \frac{y_1^{i}}{\eta + |y_1^{i}|}  \right), \quad T(\mathbf{v},\mathbf{v'}) = \frac{1}{|V|}, \label{eqn_turning}
\end{equation}
which assumes no directional  persistence \cite{Othmer:1988:MDB}.

\item Since the slime layer is very thin, typically $\sim10\mu$m, we restrict cell
movement to two dimensions.

\item Each cell divides every  $ 2$ h and is  replaced by two identical
daughter cells of age $A=0$.
\end{itemize}

We assume that cells secrete attractant at a constant rate $\gamma$ and that it is
degraded by a first-order process. The resulting evolution equation for the
attractant  is
\begin{equation}
\frac{\partial S}{\partial t}  =  D_s \triangle S + \gamma
\sum_{n=1}^{N} \delta(\mathbf{x}-\mathbf{x}^i)-\mu S
\label{eqn_attr}
\end{equation}
For simplicity, we also restrict reaction and diffusion of the attractant to two
space dimensions, which is justified as follows. Since no attractant
is added to the substrate initially, which is much thicker than the slime layer, we
assume that the attractant level is always zero in the substrate. We further assume
that the flux of the attractant at the interface of the two layers is linear in
the difference of its concentration between the two layers. Thus the loss of
attractant due to diffusion to the agar can be modeled as a linear degradation,
and the degradation constant $\mu$ in (\ref{eqn_attr}) reflects the natural
degradation rate and the flux to the substrate.

In the numerical investigations described below,  (\ref{eqn_attr}) is
solved on a square domain using the ADI method with no-flux boundary conditions,
while cells move off-grid. For each time step $\Delta t$ ($\ll $ mean run time),
(\ref{internalS}), (\ref{internalE}) are integrated for each cell and the velocity
and position are updated by Monte Carlo simulation.  Transfer of variables to
and from the grid is done using bilinear interpolating operators. A detailed
description of the numerical scheme is given in Appendix A of
\cite{Xue:2009:MMT}.

\subsection*{Radial streams result from an instability of the uniform cell  distribution}

\begin{figure}[!ht]
\begin{center}
\includegraphics[width=4in]{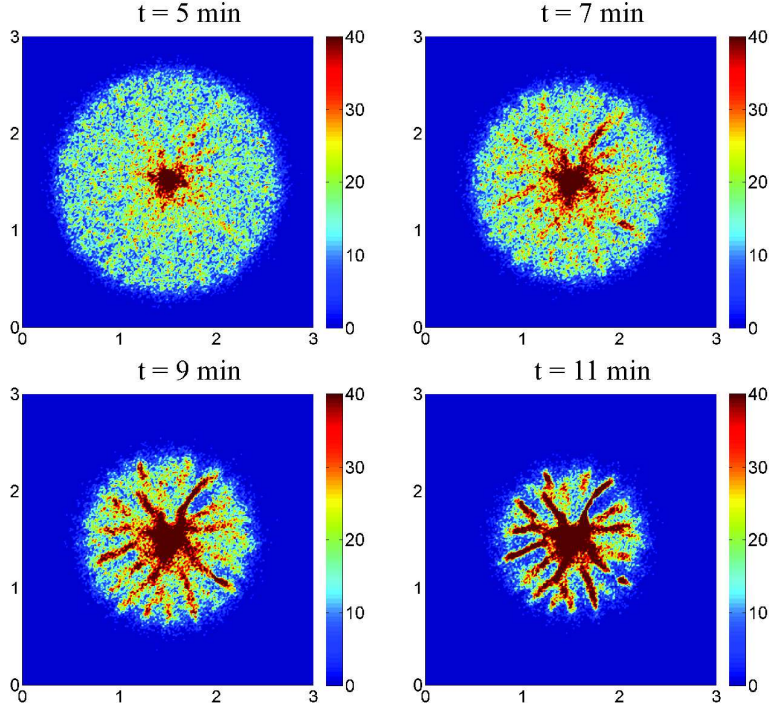}
\end{center}
\caption{
{\bf Simulated radial streams.}
 The
cell density profile is in units of $10^3$/cm$^2$. Parameters used:
$\bar{s}=20 \mu$m/s, $\lambda_0=1$/s, $D_c=9\times 10^{-6}$cm$^2$/s,
 $\mu=10^-3$/s, $L=3$cm, $t_a=5$s, $\eta=5\times 10^{-5}$, and the secretion
rate of the attractant is $6\times10^{-17}$mol/s per cell. }
\label{fig3}
\end{figure}

Radial streams appear after several hours of bacterial growth, and before their
emergence, the cell density is uniform in the colony, except at the inoculation
site.  At this stage the attractant concentration can be approximated by a
cone-like profile centered at that site.  Here we show that starting from this
initial condition, the mechanism introduced above can explain radial stream
formation. In the numerical investigations below  we assume that $t_e=0$ and
$G(S)=S$ for simplicity. Therefore,
\begin{eqnarray*}
&\displaystyle\frac{dy_2^{i}}{dt}=\frac{S(\mathbf{x},t)-y_2^{i}}{t_a}, & \vspace*{0.2cm}\\
&y_1^i=S(\mathbf{x},t)-y_2^i.&
\end{eqnarray*}
We specify an initial attractant gradient of $4\times 10^{-3}\mu$M/cm in a disk
of radius 1.5 cm, centered at the center of the domain, with zero attractant at
the boundary of the disk.  For compatibility with later computations on a
growing disk, we initially distribute $10^4$ cells/cm$^2$ randomly within
the disk. (If cells are initially distributed throughout the square domain cells
near the four corners, outside the influence of the initial gradient, aggregate
into spots, as is observed in  \textit{Escherichia coli}  as well \cite{Xue:2009:MMT}.)  Figure
\ref{fig3} shows how this distribution evolves into radial streams
that terminate in a high-density region at the center, as expected. One can
understand the breakup into streams as follows.

Whether or not there is a macroscopic attractant gradient, cells bias their run
lengths in response to the local concentration and the changes they measure via
the perceived Lagrangian derivative of attractant along their trajectory.  In
this situation, the small local  variations in cell density produce local
variations in attractant to which the cells respond.  In the absence of a
macroscopic gradient, an initially-uniform cell density evolves into a high cell
density network, which in turn breaks into aggregates, and then nearby
aggregates may merge (\cite{Xue:2009:MMT}, Figure 4.4), as is also observed
experimentally in {\em E. coli} (\cite{Budrene:1991:CPF}. If we describe the
cell motion by a 1-D velocity jump process, a linear stability analysis of the
corresponding continuum equations predicts that the uniform distribution is
unstable, and breaks up into a well-defined spatial pattern (\cite{Xue:2009:MMT}),
Figure 4.2, 4.3). Numerical solutions of the nonlinear equations confirm this,
and experiments in which the grid size is varied show that the results are
independent of the grid, given that it is fine enough
\cite{Xue:2009:MMT}.

In the presence of a macroscopic gradient a similar analysis, taken along a 1D
circular cross-section of the 2D aggregation field, predicts the breakup of the
uniform distribution, but in this situation the 2D pattern of local aggregations
is aligned in the direction of the macroscopic gradient. This is demonstrated in
a numerical experiment in which cells are placed on a cylindrical surface with
constant attractant gradient. Thus the
experimentally-observed radial streams shown in Figure \ref{fig1} and
the theoretically-predicted ones shown in Figure \ref{fig3} can be
understood as the result of (i) a linear instability of the uniform cell
density, and (ii) the nonlinear evolution of the growing mode, with growth
oriented by the initial macroscopic gradient of attractant.

\subsection*{Spiral streams result from a surface-induced swimming bias}

In most experiments the radial streams that arise initially rapidly evolve into
spiral streams, and importantly, these spirals always wind counter-clockwise
when viewed from above. The invariance of the chirality of these spirals
indicates that there are other forces that act either on individual cells or on
the fluid in the slime layer, and that initial conditions play no significant
role.  One possible explanation, which we show later can account for the
observed chirality, stems from observations of the swimming behavior of  \textit{Escherichia coli} 
in bulk solution and near surfaces.  When far from the boundary of a container,
 \textit{Escherichia coli}   executes the standard run and tumble sequence, with more or less
straight runs interrupted by a tumbling phase in which a new, essentially random
direction is chosen.  (There is a slight tendency to continue in the previous
direction). However, observations of cell tracks near a surface show that cells
exhibit a persistent tendency to swim clockwise when viewed from above
\cite{Frymier:1995:TDT, DiLuzio:2005:ECS, Lauga:2006:SCM}.

Since the cells are small, the Reynolds number based on the cell length is very
small ($ \mathcal{O}(10^{-5}$)), and thus inertial effects are negligible, and
the motion of a cell is both force- and torque-free. Since the flagellar bundle
rotates counter-clockwise during a run, when viewed from behind, the cell body
must rotate clockwise. When a cell is swimming near a surface, the part of the
cell body closer to the surface experiences a greater drag force due to the
interaction of the boundary layer surrounding the cell with that at the immobile
substrate surface.  Suppose that the Cartesian frame has the x and y axes in the
substrate plane and that z measures distance into the fluid.  When a cell runs
parallel to the surface in the y direction and the cell body rotates CW, the
cell body experiences a net force in the x direction due to the asymmetry in the
drag force. Since the flagellar bundle rotates CCW, a net force with the
opposite direction acts on the flagella, and these two forces form a couple that
produces the swimming bias of the cell. (Since the entire cell is also
torque-free, there is a counteracting viscous couple that opposes the rotation,
and there is no angular acceleration.) The closer the cell is to the surface,
the smaller is the radius of curvature and the slower the cell speed. Because of
the bias, cells that are once near the surface tend to remain near the surface,
which increases the possibility of attachment.  (In the case of {\em Proteus}
this may facilitate the swimmer-to-swarmer transition, but this is not
established.)  Resistive force theory has been used to derive quantitative
approximations for the radius of curvature as a function of the distance of the
cell from the surface and other cell-level dimensions, treating the cell body as
a sphere and the flagellar bundle as a single rigid helix \cite{Lauga:2006:SCM}.
Cell speed has been shown to first increase and then decrease with increasing
viscosity of linear-polymer solutions when cells are far from a surface
\cite{Magariyama:2002:MEI}, but how viscosity changes the bias close to a
surface is not known.

\begin{figure}[!ht]
\begin{center}
\includegraphics[width=4in]{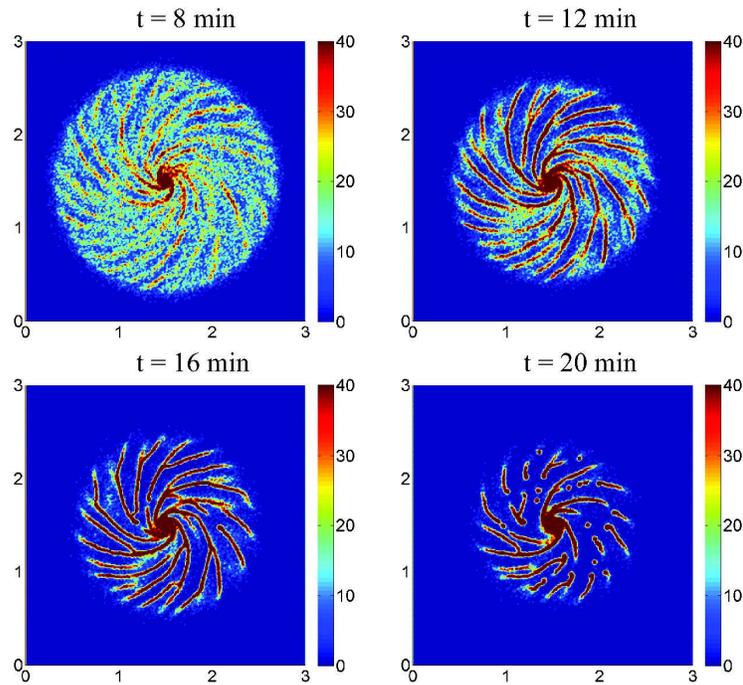}
\end{center}
\caption{
{\bf Simulated spiral streams in a disk using a swimming bias of $\varepsilon_b=0.04\pi$.} The initial
attractant gradient is $4\times 10^{-3}\mu$M/cm, centered as before, and all
other parameters are as used for the results in \protect Figure \ref{fig3}.}
\label{fig4}
\end{figure}

The question we investigate here is whether the
microscopic swimming bias of single bacteria can produce the macroscopic spiral
stream formation with the correct chirality.  We cannot apply the above theory
rigorously, since that would involve solving the Stokes problem for each cell,
using variable heights from the surface. Instead, we  introduce a constant
bias of each cell during the runs, {\em i.e.},
\begin{equation*}
\frac{d{\bf v^i}}{dt} = \varepsilon_b \frac{\bf v_i}{|{\bf
v_i}|}\times {\bf k}
\end{equation*}
where ${\bf k}$ is the normal vector to the surface, and $\varepsilon_b>0$
measures the magnitude of the bias in the direction of swimming.

\begin{figure}[!ht]
\begin{center}
\includegraphics[width=4in]{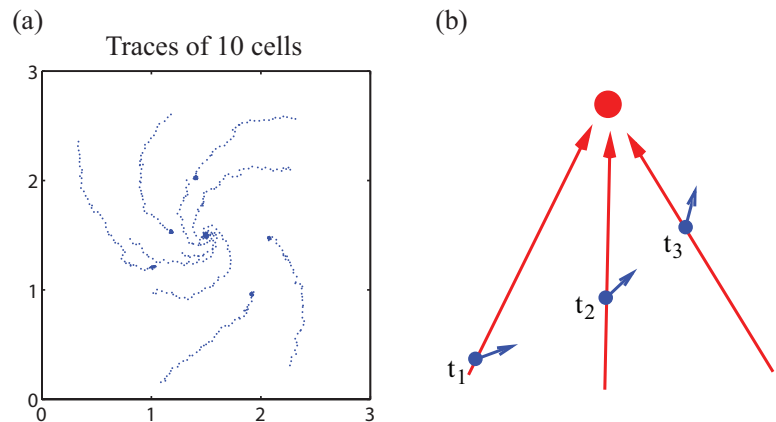}
\end{center}
\caption{(a)
The positions of 10 randomly chosen cells, each position recorded every 30 sec
by a blue dot. (b) schematics of cell movement. }
\label{fig5}
\end{figure}

Figure \ref{fig4} shows the evolution of the cell density using a bias of
$\varepsilon=0.04\pi$, which is chosen so that a cell traverses a complete
circle in 50 secs, but the results are insensitive to this choice.  The
simulations show that the initially-uniform cell density evolves into spiral
streams after a few minutes and by 12 minutes the majority of the cells have
joined one of the spiral arms.  The spiral streams persist for some time and
eventually break into necklaces of aggregates which actively move towards the
center of the domain.

Figure \ref{fig5}a illustrates the
positions of 10 randomly chosen cells every 30 seconds, and Figure
\ref{fig5}b illustrates how to understand the macroscopic
chirality based on the swimming bias of individual cells. 
At $t=t_1$ the blue
cell detects a signal gradient (red arrow) roughly in the 1 o'clock direction,
and on average it swims up the gradient longer than down the gradient. Because
of the clockwise swimming bias, the average drift is in the direction of the
blue arrow.  At $t=t_2$ it arrives at the place and `realizes' that the signal
gradient is roughly in the 12 o'clock direction, and a similar argument leads to
the average net velocity at that spot.  As a result of these competing
influences, the cell gradually make its way to the source of attractant (the red
dot) in a counterclockwise fashion. Certainly the pitch of the spirals is
related to the swimming bias, but we have not determined the precise
relationship. The spiral movement has also been explained mathematically in
\cite{Xue:2009:MMT}, where the macroscopic chemotaxis equation is derived from the hybrid
model in the presence of an external force, under the shallow-signal-gradient
assumption.  When the swimming bias is constant, the analysis shows that this
bias leads to an additional taxis-like flux orthogonal to the signal gradient.

According to the foregoing explanation, one expects spirals in the opposite
direction when experiments are performed with the petri plate upside-down and
patterns are viewed from the top, since in this case the relative position of
the matrix and slime is inverted and cells are swimming under the surface. This
prediction has been confirmed experimentally, and the conclusion is that the
interaction between the cell and the liquid-gel surface is the crucial factor
that determines the genesis and structure of the spirals.

\subsection*{Pattern formation on a growing disk}
\label{sec_integratedpicture}

\begin{figure}[!ht]
\begin{center}
\includegraphics[width=4in]{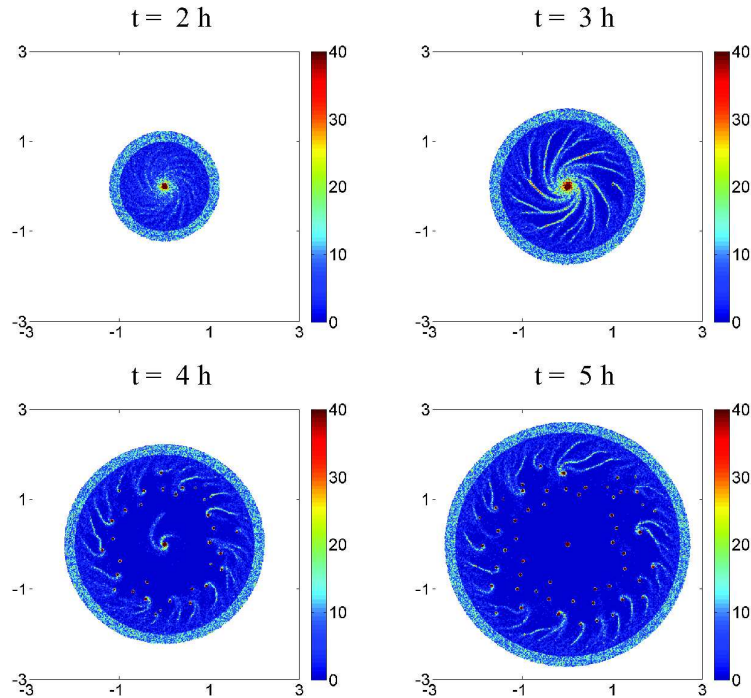}
\end{center}
 \caption{ 
{\bf Streams in a growing colony.}  $\rho_0=10^4 \mbox{cells/cm}^2$,
 $\varepsilon_b=0.04\pi$, Other parameters used
 are the same as in Figure \ref{fig3}.} \label{fig6}
 \end{figure}

\begin{figure}[!ht]
\begin{center}
\includegraphics[width=4in]{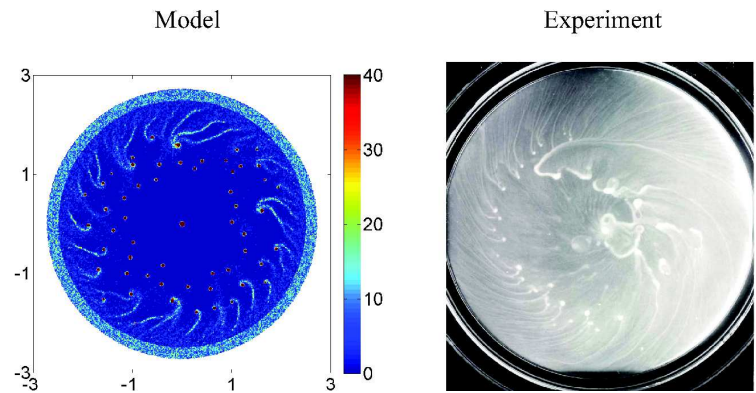}
\end{center}
\caption{ {\bf A comparison of predicted and observed spatial
patterns.} Parameters used are the same as in Figure \ref{fig3}.}
\label{fig7}
\end{figure}

From the foregoing simulations we conclude that when the swimming bias is
incorporated, the
hybrid model correctly predicts the emergence of streams and their evolution
into spirals of the correct chirality for experimentally-reasonable initial cell
densities and attractant concentration. Next we take one more step toward a complete model by
incorporating growth of the patterning domain. As we indicated earlier, the
biology of swimmer/swarmer differentiation and the biophysics of movement at the
leading edge are poorly understood. Consequently, we here regard the advancing
front as a source of swimmer cells and prescribe a constant expansion rate as observed in 
experiments (Figure \ref{fig_front}). The results of one 
computational experiment are shown in Figure \ref{fig6}, in which
the colony expands outward at a speed of $0.5$cm/h (as in Figure \ref{fig_front} after the initial lag phase), 
and the cells added in this
process are swimmer cells. One sees that the early dynamics when the disk is
small are similar to the results in Figure \ref{fig4} on a fixed disk, but as
the disk continues to grow the inner structure develops into numerous isolated
islands, while the structure near the boundary exhibits the spirals.  The
juxtaposition in Figure \ref{fig7} of the numerical simulation of the pattern at
5 hours and the experimental results shown in Figure \ref{fig1} shows
surprisingly good agreement, despite the simplicity of the model. This suggests
that the essential mechanisms in the pattern formation have been identified, but
others are certainly involved, since the experimental results show additional
structure in the center of the disk that the current model does not replicate.

\section*{Discussion}

\label{sec_discussion}

New experimental results reported here show that swimmer cells in the center of
the colony stream inward toward the inoculation site, and form a number of
complex patterns, including radial and spiral streams in an early stage, and
rings and traveling trains in later stages. These experiments suggest that
intercellular communication is involved in the spatial pattern formation. The
experiments raise many questions, including what induces the inward movement of
swimmer cells, why they move in streams, why radial streams quickly evolve into
spiral streams, and finally, why all the spirals wind counterclockwise. To
address these we developed a hybrid cell-based model in which we describe the
chemotactic movement of each cell individually by an independent velocity jump
process. We couple this cell-based model of chemotactic movement with
reaction-diffusion equations for the nutrient and attractant. To numerically
solve the governing equations, a Monte Carlo method is used to simulate the
velocity jump process of each cell, and an ADI method is used to solve the
reaction-diffusion equations for the extracellular chemicals.  The hybrid
cell-based model has yielded biologically-based answers to the questions raised
above. Starting with an estimate of the attractant level before the onset of the
radial streaming as the initial value, we predicted the formation of radial
streams as a result of the modulation of the local attractant concentration by
the cells. It is observed in  \textit{Escherichia coli} that 'runs' of single cells curve to the
right when cells swim near a surface, and we incorporated this swimming bias by
adding a constant angular velocity during runs of each cell.  This leads to
spiral streams with the same chirality as is observed experimentally. Finally,
by incorporating growth of the patterning domain we were able to capture some of
the salient features of the global patterns observed.

The streams and spirals reported here share similarities with those formed in
\textit{Dictyostelium discoideum}, where cells migrate towards a pacemaker, but there are significant
 differences. Firstly, the mechanism leading to aggregation is similar, in that
 in both cases the cells react chemotactically and secrete the
 attractant. However, since bacteria are small, they do a 'bakery search' in
 deciding how to move -- detecting the signal while moving, and constantly
 modulating their run time in response to changes in the signal. In contrast,
 \textit{D. discoideum} is large enough that it can measure gradients across it's length and
 orient and move accordingly. Thus bacteria measure temporal gradients whereas
 amoeboid cells such as \textit{D. discoideum} measure spatial gradients. In either case the
 cells respond locally by forming streams and migrate up the gradient of an
 attractant. However, spirals are less ubiquitous in \textit{D. discoideum}, and when they form
 they can be of either handedness, whereas in \textit{P. mirabilis}, only spirals wound
 counterclockwise when viewed from above have been observed, which emphasizes
 the importance of the influence of the cell-substrate interaction when cells
 swim close to the surface.  Experiments in which the patterning occurs in an
 inverted petri dish lead to spirals with an opposite handedness when viewed
 from above, which further support our explanation.  Our results imply that the
 spatial patterns observed in \textit{P. mirabilis} can be explained by the chemotactic
 behavior of swimmer cells, and suggest that differentiation and
 de-differentiation of the cells at the leading edge does not play a critical
 role in patterning, but rather serves to expand the colony under appropriate
 conditions. A future objective is to incorporate a better description of the
 dynamics at the leading edge when more biological information is available.

The spatial patterns reported here are also different from those observed in
other bacteria such as  \textit{Escherichia coli}  or \textit{Bacillus subtilis}.  In the latter,
fractal and spiral growth patterns have been observed
\cite{Ben-Jacob:1994:GMC,Ben-Jacob:1997:CCM}, and these patterns form primarily
at the leading edge of the growing colony. There cell motility plays a lesser
role and the limited diffusion of nutrient plays an important role in the
pattern formation.

Of course the experimental reality is more complicated than that which our model
describes. For instance, the nutrient composition is very complex and nutrient
depletion may occur at a later stages such as during train formation. Further,
cells may become non-motile for various reasons, and these factors may play a role
in the stabilization of the ring patterns. Another important issue is the
hydrodynamic interaction of the swimmer cells with fluid in the slime
layer. When cell density is low and cells are well separated we can approximate
their movement by independent velocity jump processes plus a swimming bias, but
when the cell density is high the cell movement is correlated through the
hydrodynamic interactions and this must be taken into account. This hydrodynamic
interaction may be an important factor in the formation of the trains observed
in experiments.

The individual cell behavior, including the swimming bias, has been embedded in
a continuum chemotaxis equation derived by analyzing the diffusion limit of a
transport equation based on the velocity jump process \cite{Xue:2009:MMT}. The
resulting equation is based on the assumption that the signal gradient is
shallow and the predicted macroscopic velocity in this regime is linear in the
signal gradient.  A novel feature of the result is that the swimming bias at the
individual cell level gives rise to an additional taxis term orthogonal to the
signal gradient in this equation. However in the simulations of the patterns
presented here we observe steep signal gradients near the core of the patterns
and within the streams, and therefore in these regimes the assumptions
underlying the continuum chemotaxis model are not valid. A statistical analysis
of cell trajectories in the results from the cell-based model reveals saturation
in the macroscopic velocity and a decreasing diffusion constant as the signal
gradient grows, which suggests that in the limiting case of large gradients, the
macroscopic equation for cell density will simply be a transport equation with
velocity depending on the signal gradient.

\section*{Materials and Methods}

\subsection*{Chemotaxis analysis of swimmer cells}
\label{appendix_taxis}
Positive chemotaxis toward each of the common 20 amino acids was
tested using the drop assay. Each amino acid was tested at the following concentrations: .1M,
10mM , 1mM, l0 $\mu$M, and 1 $\mu$M.

\begin{table}[!hb]

\centering
\begin{tabular}{|c|c|c|c|c|c|}
\hline
 &   .1M   &   10mM   &   1mM   &  10 $\mu$m &   l $\mu$m \\
\hline Ala  & +   &     +   &    -    &   -   &    - \\
 Arg  & -   &     -   &    -    &   -   &    -\\
 Asn  & -   &     +   &    -    &   -   &    -\\
 Asp  & +   &     -   &    +    &   +   &    +\\
 Cys  & -   &     -   &    +    &   +   &    -\\
 Glu  & +   &     +   &    -    &   -   &    -\\
 GIn  & -   &     -   &    -    &   -   &    -\\
 Gly  & +   &     +   &    -    &   -   &    -\\
 His  & +   &     +   &    -    &   -   &    -\\
 Ile  & -   &     -   &    -    &   -   &    -\\
 Leu  & -   &     -   &    -    &   -   &    -\\
 Lys  & -   &     -   &    -    &   -   &    -\\
 Met  & +   &     +   &    -    &   -   &    -\\
 Phe  & -   &     +   &    +    &   -   &    -\\
 Pro  & -   &     -   &    -    &   -   &    -\\
 Ser  & +   &     +   &    +    &   +   &    -\\
 Thr  & +   &     -   &    -    &   -   &    -\\
 Trp  & -   &     -   &    -    &   -   &    -\\
 Val  & -   &     -   &    -    &   -   &    -\\
 \hline
\end{tabular}
\caption{ Amino acid drop assay. \textit{Proteus} cells were collected from
the inner area of a growing colony, approximately 1 hr before a
projected onset of a streaming phase. Microscopic examination
revealed that 90\% of cells were 1 to 2 cell length. Cells were
resuspended in a minimal growth medium to the OD=.1 to .15 (similar
results were obtained with the cells grown in a liquid culture) Drop
Assay. 500 $\mu$L minimal growth medium, 200 $\mu$L of cell culture
(OD=.l to .15), and 240 $\mu$L of 1\% Methyl cellulose were combined
in a l0x35 mm culture dish and mixed until a homogenous state. 4
$\mu$L of a respective amino acid solution was added to the center.
Cell density distribution in the dish was analyzed after 20-25
minutes. Addition of H$_2$O was used as a control. Increase in the
cell density in the center indicates that a respective amino acid is
an attractant. }
 
\end{table}

\subsection*{Chemotaxis on semi-solid agar}

Chemotaxis of swimmer cells towards single amino acids was tested
using 0.3\% agar plates with different thickness of substrate
layer(10 and 20 ml). Each amino acid was used in concentrations
varying from 0.25mM to 7.5mM in both thicknesses of agar. The plates
were point inoculated and placed in a humid chamber at room
temperature for at least 20 hrs. Bacteria growing on 10 and 20 ml
plates with 0.00lM of Aspartate, Methionine and Serine formed dense
moving outer ring which we interpret as a chemotactic ring. Bacteria
grown on all remaining amino acids produced colonies with the higher
density at the point of inoculation and homogeneous cell
distribution in the rest of the colony.

\subsection*{Numerical algorithm}

\begin{figure}[!ht]
\begin{center}
\includegraphics[width=2in]{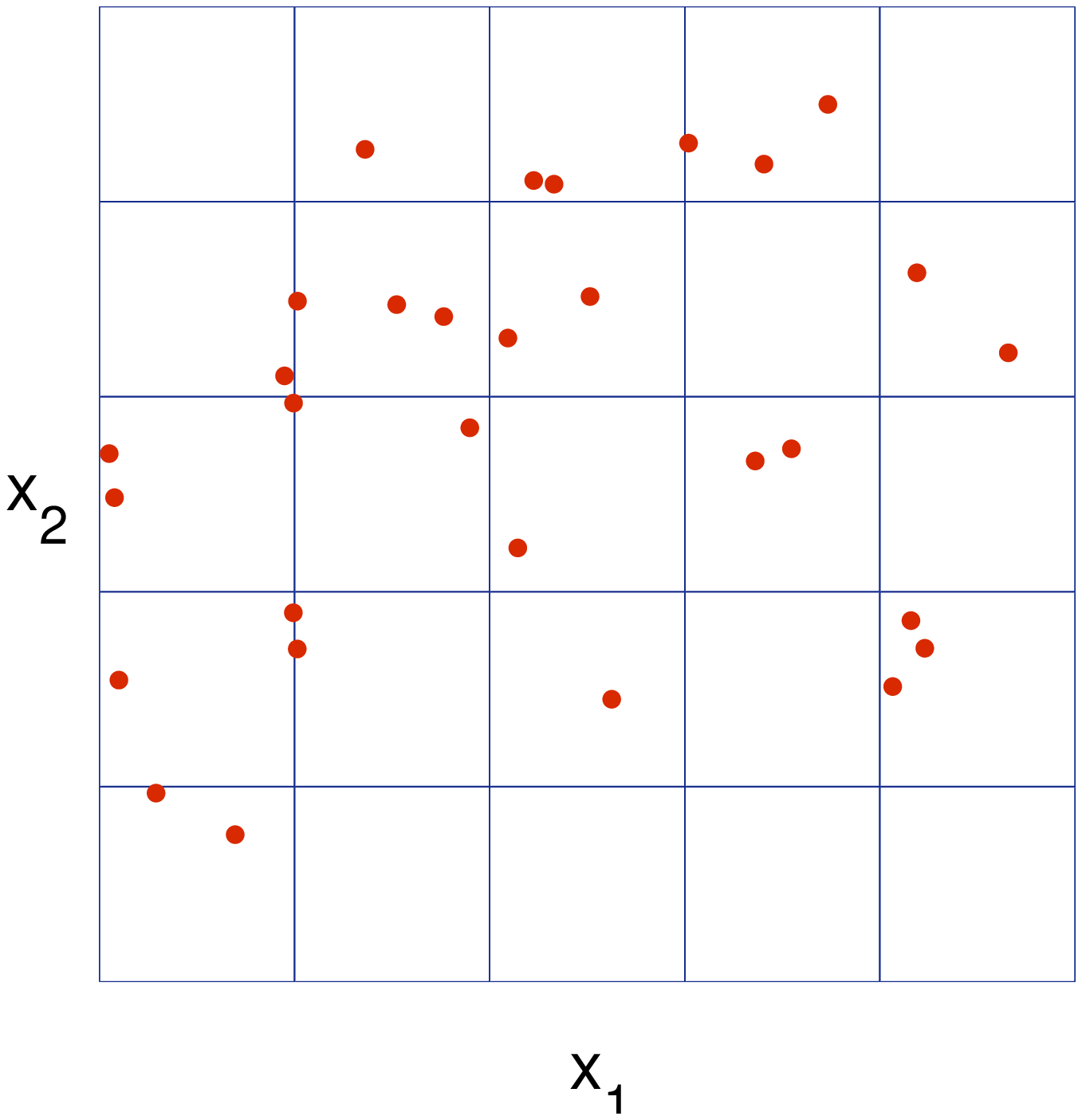}\hspace*{1cm}
\includegraphics[width=2in]{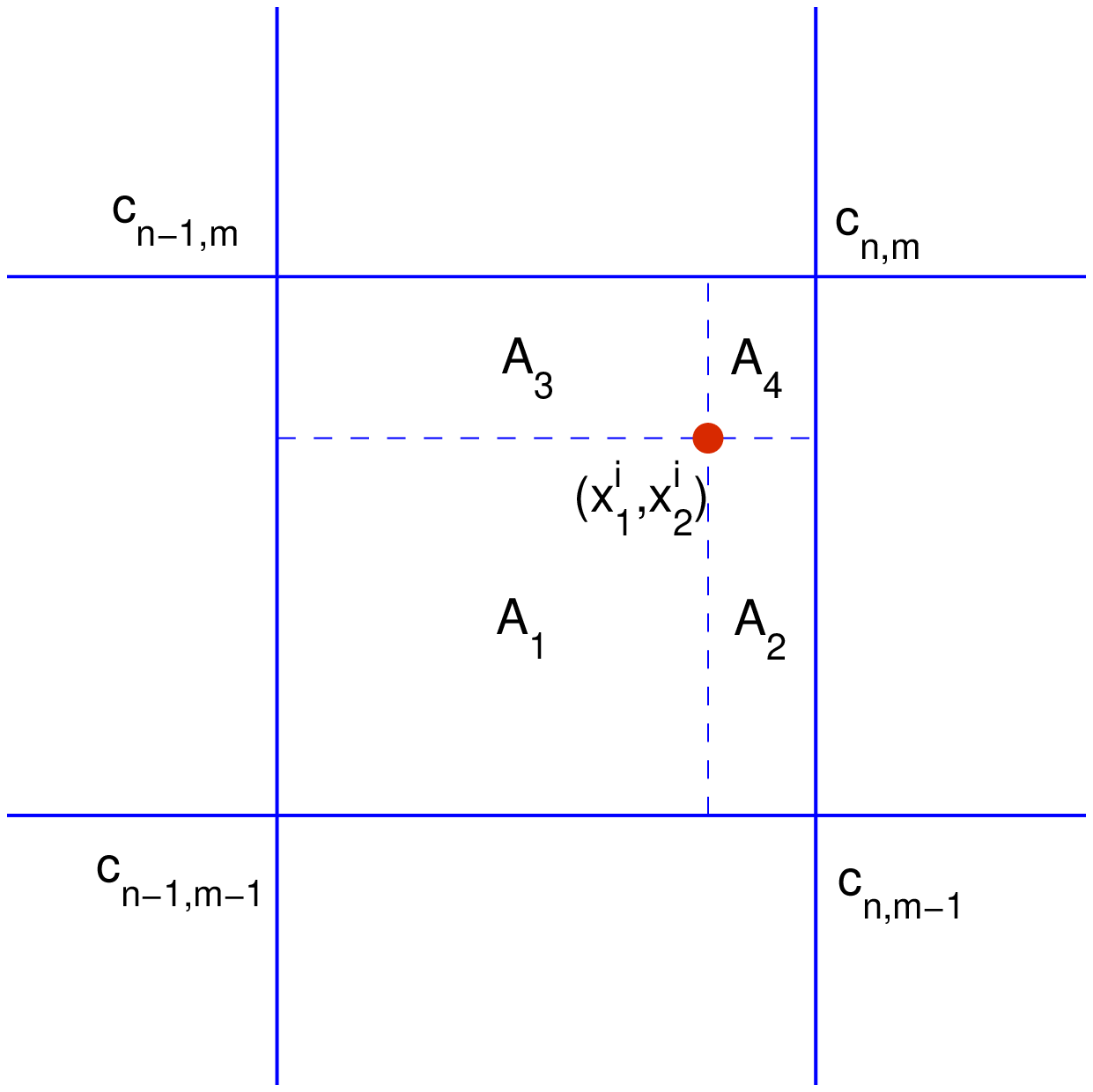}
\end{center}
\caption{
{\bf Numerical algorithm.} Left: a schematic figure of the domains. The
reaction-diffusion equations are solved on the grid, while the
cells can move around the whole domain. Right: the area fractions
used in defining the interpolators (\ref{eqn_tgc}, \ref{eqn_tcg}).
} \label{fig_interp}
\end{figure}

In the implementation of the cell-based model, cell motion is
simulated by a standard Monte Carlo method in the whole domain,
while the equations for extracellular chemicals are solved by an
alternating direction method  on a set of
rectangular grid points . In this
appendix, we present the numerical algorithm in a two-dimensional
domain with only one chemical -- the attractant -- involved . Each cell is described by its
position $(x^i_1,x^i_2)$, internal variables $(y^i_1, y^i_2)$,
direction of movement $\theta^i$ and age $T^i $ (the superscript
$i$ is the index of the cell). Concentration of the attractant is
described by a discrete function defined on the grid for the
finite difference method (Figure \ref{fig_interp}, left). We
denote the time step by $k$, the space steps by $h_1$ and $h_2$.

Since two components of the model live in different spaces, two
interpolating operators are needed in the algorithm.
$\mathcal{T}_{gc}$ is used to evaluate the attractant
concentration that a cell senses. For a cell at $(x^i_1,x^i_2)$,
inside the square with vertex indices $(n-1,m-1)$, $(n,m-1)$,
$(n-1,m)$ and $(n,m)$, $\mathcal{T}_{gc}(x^i_1,x^i_2)$ is defined
by the bi-linear function:
\begin{equation}
\mathcal{T}_{gc}(x^i_1,x^i_2) = \frac{A_4}{A} S_{n-1,m-1}+
\frac{A_3}{A} S_{n,m-1}+ \frac{A_2}{A} S_{n-1,m}+ \frac{A_1}{A}
S_{n,m}\label{eqn_tgc}
\end{equation}
where $A=h_1 h_2$ and $A_j, j=1,2,3,4$ are the area fractions
(Figure \ref{fig_interp}, right). On the other hand, the
attractant secreted by  cells is interpolated as increments at the
grid points by $\mathcal{T}_{cg}$. Suppose during one time step
$k$, a cell staying at $(x^i_1,x^i_2)$ secretes $\Delta $ amount
of attractant, we then interpolate the increment of the attractant
concentration at the neighboring grid points as follows:
\begin{equation}
 \mathcal{T}_{cg}(\mathbf{x}^i;p,q)  = \left\{
\begin{array}{cl}
\vspace*{0.1cm}\frac{A_4\Delta}{A^2}, &\quad (p,q)=(n-1,m-1);\\
\vspace*{0.1cm}
\frac{A_3\Delta}{A^2}, &\quad (p,q)=(n,m-1 ); \\
\vspace*{0.1cm}\frac{A_2\Delta}{A^2}, &\quad (p,q)=(n-1 ,m );   \\
\vspace*{0.1cm}\frac{A_1\Delta}{A^2}, &\quad (p,q)=(n ,m ); \\
0, &\quad  \mbox{ otherwise.}
\end{array}
\right.\label{eqn_tcg}
\end{equation}

We consider here a periodic boundary condition. The detailed
computing procedure is summarized as follows.

\begin{enumerate}

\item[\bf S1] Initialization.

\begin{enumerate}

\item Initialize the chemical fields.

\item Initialize the list of swimmer cells. Each cell is put in
the domain with random position, moving direction and age.
$\mathbf{y}^i$ is set to be $0$.

\end{enumerate}

\item[\bf S2] For time step $l$ ($=1$ initially), update the data
of each cell.

\begin{enumerate}
\item Determine the direction of movement $\theta^i$ by the turning kernel. \\
i) Generate a random number $r\in U[0,1]$; \\
ii) If $r<1-e^{-\lambda^i k}$, update $\theta^i$ with a new random
direction.

\item $(x_1^i, x_2^i)_l\longleftarrow (x^i_1, x^i_2)_{l-1}+(sk\cos
\theta^i,sk\sin \theta^i)$. Apply periodic boundary condition to
make sure $(x_1^i, x_2^i)$ inside the domain,

\item $(T^i)_l\longleftarrow (T^i)_{l-1}+k$. If $(T^i)_l \geq 2$
hours, then divide the cell into two daughter cells. This step is
only considered when cell growth is considered.

\item Update $(y^i_1, y^i_2)$ by the equations for the internal dynamics. \\
i) Determine the attractant concentration before the cell moves $(S^i)_{l-1}$ and after the cell moves $(S^i)_{l}$ by using the interpolating operator $\mathcal{T}_{gc}$. \\
ii) Estimate the attractant level during the movement by $S^i(t) = (S^i)_{l-1} \frac{t-lk}{k} + (S^i)_{l} \frac{lk+k-t}{k}$ and integrate equation for $y^i_2$ to get $(y^i_2)_l$. \\
iii) $(y^i_1)_l \longleftarrow G(S)-(y^i_2)_l$.
\end{enumerate}

\item[\bf S3] Compute the source term of the attractant $
f^{l-\frac{1}{2}}$ due to the secretion by the cells using the
interpolator $\mathcal{T}_{cg}$
$$f^{l-\frac{1}{2}}_{p,q}=\sum_i (\mathcal{T}_{cg}((\mathbf{x}^i)_{l-\frac{1}{2}};p,q)),$$
where $\Delta=\gamma k$.

\item[\bf S4] Apply the alternating direction implicit method to
the equation of the attractant: {\small
\begin{eqnarray*}
\frac{S^{l-1/2}_{p,q}-S^{l-1}_{p,q}}{k/2} &=& D_s \frac{S_{p+1,q}^{l-1/2} - 2S_{p,q}^{l-1/2}+ S_{p-1,q}^{l-1/2}}{h_x^2}  \\
&&\quad +
 D_s\frac{S_{p,q+1}^{l-1} - 2S_{p,q}^{l-1}+ S_{p,q-1}^{l-1}}{h_x^2}
- \gamma \frac{S_{p,q}^{l-1}+S_{p,q}^{l-1/2}}{2} + f_{p,q}^{l-\frac{1}{2}}, \\
\vspace*{0.3cm}
\frac{S^{l}_{p,q}-S^{l-1/2}_{p,q}}{k/2} &=& D_s \frac{S_{p+1,q}^{l-1/2} - 2S_{p,q}^{l-1/2}+ S_{p-1,q}^{l-1/2}}{h_x^2} \\
&&\quad + D_s \frac{S_{p,q+1}^{l } - 2S_{p,q}^{l }+ S_{p,q-1}^{l
}}{h_x^2}- \gamma \frac{S_{p,q}^{l-1/2}+S_{p,q}^{l }}{2} +
f_{p,q}^{l-\frac{1}{2}}.
\end{eqnarray*}
} For the boundary grid points, use the periodic scheme.

\item[\bf S5] $l\longleftarrow l+1$. If $lk\leq T_0$, repeat {\bf
S2-S4}; otherwise, return.
\end{enumerate}

\section*{Acknowledgments}
CX is supported by the Mathematical Biosciences Institute under the US NSF Award
0635561. HGO is supported by NIH grant GM 29123, NSF Grant DMS 0817529 and
the University of Minnesota Supercomputing Institute.


\end{document}